\documentclass{webofc}
\usepackage[varg]{txfonts}  
\usepackage{graphics}
\usepackage{amsmath,amssymb}

\usepackage{epsfig}
\usepackage{hyperref}
\usepackage{amsopn}
\usepackage{float}
\usepackage{tikz}
\usepackage{graphicx}
\usepackage{ulem}
\usepackage{cancel}
\usepackage{color}

\newcommand{\be}{\begin{equation}}
\newcommand{\ee}{\end{equation}}
\newcommand{\bea}{\begin{eqnarray}}
\newcommand{\eea}{\end{eqnarray}}

\newcommand{\GeV}{\:\text{GeV}}

\begin{document}

\title{Upper limit on first-order electroweak phase transition strength}

\author{ Shehu AbdusSalam\inst{1}\thanks{\emph{Emails:} abdussalam@sbu.ac.ir, l\_kalhor@sbu.ac.ir, mj\_kazemi@sbu.ac.ir} \and Layla Kalhor\inst{1} \and Mohammad Javad Kazemi\inst{1}}
\institute{Department of Physics, Shahid Beheshti University, Tehran, Islamic Republic of Iran} 

\abstract{For a cosmological first-order electroweak phase transition, requiring no sphaleron washout
  of baryon number violating processes leads to a lower bound on the strength of the transition. 
  The velocity of the boundary between the phases, the so-called bubble wall, can become ultra-relativistic
  if the friction due to the plasma of particles is not sufficient to retard the wall's acceleration.
  This bubble ``runaway''  
  should not occur if a successful baryon asymmetry generation due to the transition is required.
  Using Boedeker-Moore criterion for bubble wall runaway, within the context of an extension of
  the standard model of particle physics with a real gauge-single scalar field, we show that a
  non runaway transition requirement puts an upper bound on the strength of the first-order 
  phase transition.}

\onecolumn
\maketitle
\section{Introduction}
The universe we observe and understand is mainly made of matter and has very little antimatter component. There
are two important shortcomings of the standard model (SM) of particle physics due to this observation -- insufficient
source for CP violation~\cite{Gavela:1993ts} and scalar sector interactions to allow for a good phase
transition~\cite{Kajantie:1995kf,Kajantie:1996mn} at early time of the cosmos. This is because the observed matter-antimatter
asymmetry could be due to strong first order EW phase transition (EWPT)~\cite{Cohen:1993nk,Morrissey:2012db}
that allows for primordial baryon number and CP violating processes~\cite{Sakharov:1967dj}. This mechanism is
called electroweak baryogenesis (EWBG) \cite{Bochkarev:1990fx,Cohen:1990py,Cohen:1990it,Turok:1990zg}. 
For a first-order EWPT, EWBG will work only if the baryon asymmetry created at the expanding bubble wall is
not washed out by sphaleron~\cite{Kuzmin:1985mm} processes inside the broken phase. For this reason, a strong enough first-order
EWPT is required for EWBG to be successful. 

Another consequence for a strong first-order EWPT is that it could generate stochastic background of
gravitational waves (GW)~\cite{Witten:1984rs,Hogan:1986qda,Kosowsky:1991ua,Kamionkowski:1993fg} which
(see, e.g.~\cite{Huang:2016cjm} for prospects studies within the context of the model we considered)
could be detected at future satellite GW interferometers~\cite{Caprini:2015zlo}.
The velocity of the expanding bubbles is one of the essential parameters that characterises the
EWBG and GW dynamics related to the transition~\cite{Steinhardt:1981ct,Apreda:2001us,Kosowsky:1991ua,Laine:1993ey,Ignatius:1993qn,
  Moore:1995si,Kamionkowski:1993fg,Caprini:2007xq,Huber:2008hg,Caprini:2009fx,Kosowsky:2001xp,Caprini:2006jb,Cline:2020jre}. 
Observable gravity waves require very strong phase transition but this leads to fast-moving bubbles. 
In general, the speed of the accelerating bubble wall can be retarded by friction due to the collisions with particles in its
surrounding plasma. This way, the speed could reach a steady state after some finite time. Otherwise, if the friction
is not sufficient, it keeps increasing towards ultra-relativistic magnitudes for very strong first-order
phase transitions~\cite{0903.4099}. The latter scenario leads to the so-called ``runaway'' of the
bubble wall for which EWBG will not work because there is not enough time for the baryon-antibaryon asymmetry to be 
generated outside the bubbles~\cite{0903.4099,Kurup:2017dzf}.

So two trends can be spotted. On one hand, for EWBG to explain the matter-antimatter asymmetry of the universe
a strong first-order EWPT is required. There is no precise quantification~\cite{Patel:2011th,Fuyuto:2014yia} for how strong
the transition must be; but conventionally a certain lower limit is usually assumed, $v_n/T_n > 1$ where $v_n$ is the vacuum
expectation value (VEV) of the SM Higgs field at the bubble nucleation temperature $T_n$. Very strong EWPT, on the other hand,
could yield stronger GW signals
but may also lead to the runaway scenario and conflict with EWBG. Many work have been done~\cite{Huang:2016cjm,
  Artymowski:2016tme,Hashino:2016xoj,Chao:2017vrq,Beniwal:2017eik,Kurup:2017dzf,Baldes:2017rcu,Angelescu:2018dkk,
  Beniwal:2018hyi,Ahriche:2018rao,Athron:2019teq} along these directions and mostly focusing on the combination of
imposing a no sphaleron washout condition $v_n/T_n > 1$ and possibility of the GW observation. In this article we address
the question: for a successful EWBG, how large can $v_n/T_n > 1$ be? This is done within the context of an inert
singlet~\cite{Burgess:2000yq,Espinosa:2011ax} extension of the SM. The model parameters can readily be found capable
to generate strong phase transitions with high bubble wall velocity. We found that requiring no runaway of
bubble walls, using the Boedeker-Moore condition~\cite{0903.4099}, puts an upper bound on how strong the EWPT could be.
The analyses were made partly using cosmoTransitions~\cite{Wainwright:2011kj} package for finding transition and nucleation
temperatures, bubble profiles, and computing the GW wave power spectra. We check some of the cosmoTransitions 
results and found that they agree with those from BSMPT (Beyond the Standard Model Phase Transitions)~\cite{Basler:2018cwe}
using the same model parameters.

Our approach adds to that in~\cite{0903.4099} as follows.  
  In \cite{0903.4099} the zero-temperature one-loop correction, i.e. the Coleman-Weinberg potential, were
not taken into consideration and a high-temperature approximation is used for thermal corrections to the effective potential. We have considered a $Z_2$ symmetric potential with the full one-loop and thermal corrections for our analyses. Further, we have considered a positive mass-term for the singlet field, contrary to the case considered in \cite{0903.4099}. In fact, depending on this sign, there are two different scenarios for EWPT. First, for the case we have considered, the singlet never attains a VEV, and there are no tree-level effects to enhance the phase transition. However, it is still possible to induce a strong electroweak phase transition via sizable one-loop zero-temperature corrections to the SM Higgs potential. In this case, the phase transition occurs purely along the Higgs direction and leads to a one-step phase transition. Secondly for a positive mass-term of the singlet field in the potential, the early universe can transit into a minimum along the singlet field direction before the EWPT occurs. The positive mass-term leads to tree-level modifications of the potential. This can make it possible for strongly first-order EWPT to happen even without requiring one-loop correction to the zero-temperature potential. Lastly, our result goes beyond the inert singlet model and could be model independent. See the results in Appendix~\ref{IRT} for the Inert Real Triplet Model.

\section{Model and analyses setup}  \label{Model}
In this section we set the model, context and notations for analysing the correlations between the requirements for
no sphaleron washout and no bubble wall runaway following the primordial electroweak phase transition. The beyond the
SM theory we consider is one with a scalar singlet $S$ added to the Higgs sector. We explore the parameter space of the model 
and identify regions where a first order transition occur. For the sample of parameter points generated we compute the critical, $T_c$,  
and nucleation, $T_n$, temperatures and the GW spectrum that could arise from the EWPT. 

The tree level potential with a $\mathbb{Z}_2$ symmetry that forbids Higgs-singlet mixing is: 
\begin{equation}
	V_{\textrm{tree}} (H,S) = -\mu_H^2|H|^2+ \lambda_H |H|^4+\lambda_{HS}|H|^2 S^2+\frac{1}{2} \mu_{S}^2 S^2+\frac{1}{4}\lambda_{S}S^4,
\end{equation}
where 
\begin{equation}
	H = \frac{1}{\sqrt{2}}
	\begin{pmatrix}
		\chi_1+i \chi_2 \\
		h + i \chi_3
	\end{pmatrix},
\end{equation}
and $\chi_{\{1,2,3\}}$ are the Goldstone bosons. So, the potential in terms of the physical Higgs $h$
and singlet scalar $S$ is: 
\begin{equation}
V_{\textrm{tree}}(h,S)=-\frac{1}{2}\mu_H^2 h^2 +\frac{1}{4}\lambda_H h^4+\frac{1}{2}\lambda_{HS} h^2 S^2+\frac{1}{2} \mu_{S}^2 S^2+\frac{1}{4}\lambda_{S}S^4.
\end{equation}

The Higgs physical mass and self coupling are fixed at $m_H=125\GeV$ 
and $\lambda_H=m_H^2/2v^2\approx 0.129$ respectively 
while its VEV at zero temperature is set to $v = 246 \GeV$. 
The physical mass of the new scalar $S$ after electroweak symmetry breaking (EWSB) is $m^2_S =\mu_S^2 + \lambda_{HS}v^2$. The
full effective potential used for our analyses consists of three parts, namely the tree level, one-loop correction, and
thermal corrections terms as presented in Appendix~\ref{Veff}. 
We consider the case of $\mu_S^2$ positive. In this case the symmetry of singlet is not
spontaneously broken and thus we address only the one dimensional potential along the $h$ direction, leading to
a one-step phase transition when the field tunnels through the energy barrier between the zero minimum
and the non-zero minimum. In addition, $\lambda_{HS}$  and $\lambda_S$ were required to be positive,  
allowing of stable minimum of the potential energy. Hence there are three inert single free parameters, 
$\mu_S$, $\lambda_S$, and $\lambda_{HS}$ which will be varied simultaneously in the ranges [1, 5000] GeV, [0.01, 2.0], [0.05, $2 \pi$] respectively. Perturbativity is controlled mostly by the singlet field quartic coupling $\lambda_S$ which we have chosen to be small for our analysis. Further, we have passed the parameter points to Lilith~\cite{Bernon:2015hsa,Kraml:2019sis} to ensure consistency with Higgs-sectors collider results. The $Z_2$-symmetry and a zero VEV for singlet field at the zero-temperature forbids Higgs-singlet mixing and thus prevents the modifications of the oblique parameters. For each parameter space point, we 
used cosmoTransitions~\cite{Wainwright:2011kj} for determining $T_c$, $T_n$ and the corresponding VEVs, $v_{c,n}$, 
of the EWPT. Only parameter
regions with $\frac{v_c}{T_c} > 1$ were accepted for further analyses. Next we address the characteristics of
the bubble wall velocity and gravitational waves spectrum that could arise due to the strong first-order
EWPT. 

In order to analyse the EWPT bubble wall velocity, $v_w$, and estimate the gravitational wave power spectrum that
could result from the inert singlet model, two other quantities need to be determined. These
are~\cite{Grojean:2006bp, Caprini:2015zlo} the ratio of released latent heat from the transition to the energy
density of the plasma background, $\alpha$, and the time scale of the phase transition, $H/\beta$.
Using the effective potential and it's derivative at nucleation temperature, $T_n$, the parameters $\alpha$
reads as \cite{Caprini:2015zlo},
\begin{equation}\label{eqn:alpha}
\alpha= \left.\frac{1}{\rho_{R}}\left[-(V_{\rm EW}-V_f)+ T_n \left(\frac{dV_{\rm EW}}{dT} - \frac{dV_f}{dT}\right)\right]\right|_{T=T_n} ,
\end{equation}
where $V_f$ is the value of the potential in the unstable vacuum  and $V_{\rm EW}$ is the value of the potential in the final vacuum. The time scale of the phase transition can be calculate from the derivative of the Euclidean action at
nucleation temperature~\cite{Caprini:2015zlo}:  
\begin{equation}\label{eqn:beta}
\frac{H}{\beta}= \left. \left[T \frac{d}{dT} \left(\frac{S_3(T)}{T} \right)\right]^{-1}\right|_{T=T_n}.
\end{equation}
The calculation of the bubble wall velocity is not within the scope of this article. For this there is need to consider
the interaction between the wall and surrounding plasma. However there are approximate expressions in terms of
$\alpha$ such as in~\cite{Steinhardt:1981ct}: 
\begin{equation} \label{bubblevel}
v_w=\frac{1/ \sqrt{3}+\sqrt{\alpha ^2+2 \alpha /3}}{1+\alpha } 
\end{equation}
which represents a lower bound on the true wall velocity~\cite{Huber:2008hg}. 
In this work, we use this approximation together with the expressions for $\alpha$ and $H/\beta$ for
calculating the GW signals produced during the phase transition. 

Depending on the bubble wall velocity there are two main regimes;  when the wall velocity is relativistic or not.    
In addition, in the relativistic regime, there are two qualitatively different scenarios. First, if bubble wall reaches a terminal velocity (non-runaway scenario), second, the bubble wall accelerates without bound (runaway scenario). In order to calculate the GW spectrum, it is important to know which of the aforementioned  scenarios apply. To this aim, the critical $\alpha$ value, $\alpha_{\infty}$, can be used to distinguish between these scenarios \cite{Caprini:2015zlo, Espinosa:2010hh},
\begin{equation}
\alpha_{\infty} \simeq \frac{30}{24 \pi^2}\frac{\sum_a c_a \Delta m_a^2}{g_* T_*^2} \simeq 0.49 \times 10^{-3} \left(\frac{v_n}{T_n}\right)^2.
\end{equation}
where $c_a = n_a/2$ $(c_a = n_a)$ and $n_a$ is the number of degrees of freedom for boson (fermion) species and $ \Delta m^2_a$ is the squared mass difference of particles between two phases. 
For non-runaway scenarios, $\alpha < \alpha_{\infty}$, the wall velocity $v_w$ remains below threshold and the available energy is transformed into fluid motion. 

Another criterion for determining whether the bubble walls runaway or reach steady speed goes back to the work 
in~\cite{Bodeker:2009qy,Bodeker:2017cim}. The pressure on the wall come from two 
sources with opposite directions. One is outwards and due to the difference in energy densities of the symmetric
and broken vacuum, $V_{sym} - V_{br}$. The other is inwards and due to the pressure $P$ from the thermal plasma of
particles surrounding the wall. For each point in the parameter space of the inert singlet model we compute the
{\it Boedeker-Moore (BM) criterion} and require that \be p_{runaway} = V_{sym} - V_{br} - P < 0, \ee
where 
\be
P \approx \sum_i \left( m_{i,br}^2 - m_{i,sym}^2 \right)
\frac{g_i T_n^2}{4\pi^2}  \tilde{J}_i \left(\frac{m_{i,sym}^2}{T_n^2}\right) \quad \textrm{ and } \quad 
\tilde{J}_i(x) = \int_0^\infty \frac{y^2 dy}{\sqrt{y^2+x}} \frac{1}{e^{\sqrt{y^2+x}}+(-1)^{F_i}}.
\ee
Here $i$ runs over the considered SM and inert singlet scalar particles, while $g_i$ and $F_i$ are the particle
multiplicity and fermion number respectively.
It should be noted that the non-runaway requirement is a weak condition for EWBG. In fact, the subsonic wall velocity, $v < 1/\sqrt{3}$, is a stronger requirement since it leads to efficient diffusion of particle asymmetries in front of the bubble wall~\cite{subsonic}.

\section{Result and outlook}
\begin{figure}[H]
{\begin{tikzpicture}
\begin{centering}
\draw (-0,0) node[above right]{\includegraphics[width=0.8\linewidth]{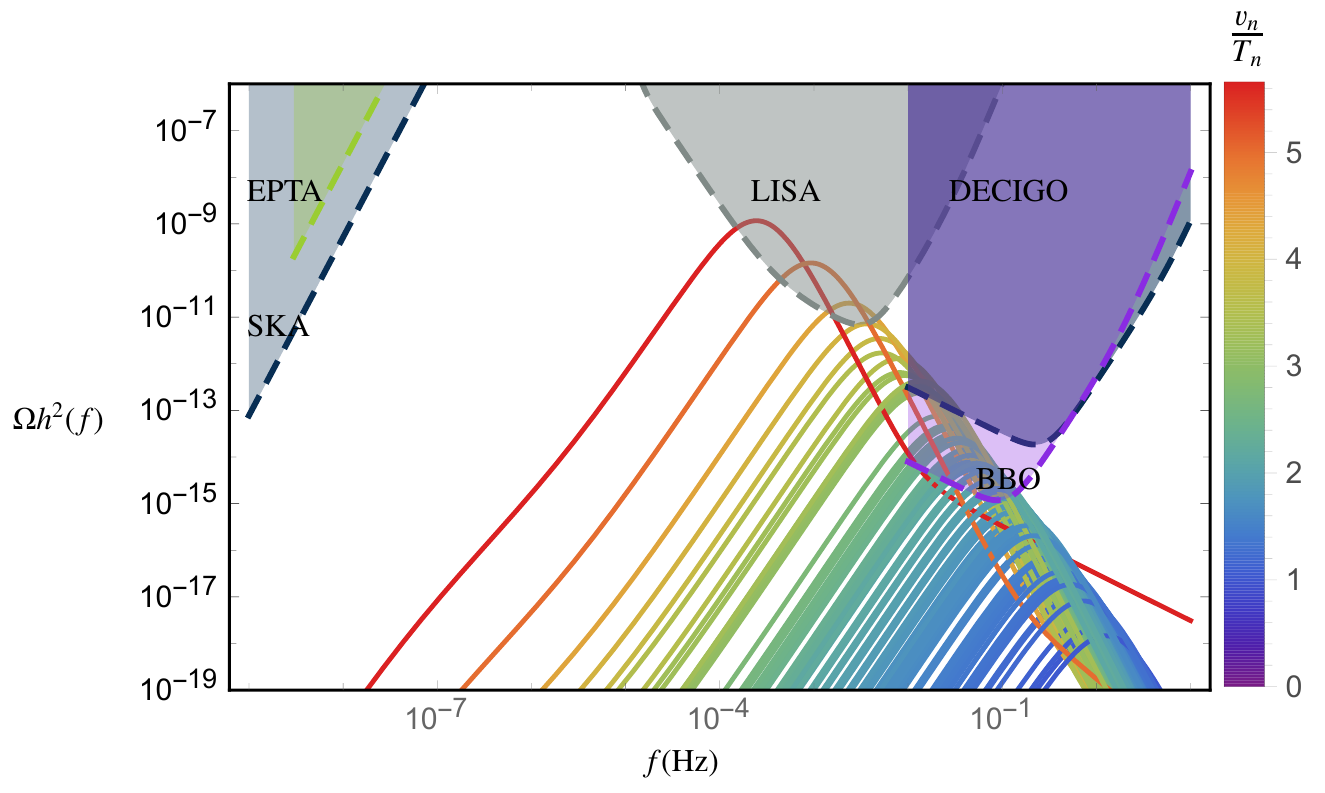}};
\end{centering}
\end{tikzpicture}
}
\caption{\label{GW-Spectrum}  Spectra of GWs from the electroweak phase transition for randomly
  sampled examples from the coloured points in figure \ref{DetectableRegion}, i.e. the points with
  strong first-order EWPT. The sensitivity region for prospective GW detectors such as LISA, BBO and
  DECIGO are also shown. It can be seen that the intensity of GW signal increases with the strength
  of the phase transition, i.e. $\frac{v_n}{T_n}$. For comparison we also show the sensitivity
  regions for SKA and EPTA detectors which cannot probe any part of the inert singlet parameter space.} 
\end{figure}

The scatter plots on ($\mu_S$, $\lambda_{HS}$) plane in figure~\ref{DetectableRegion}(left) shows a sample of the inert
singlet model parameter points indicating regions where the strong phase transition could lead to
GW accessible to promising future GW detectors, specifically LISA, DECIGO and BBO~\cite{Yagi:2011wg}. Typically,
the magnitude of the GW signal increases with the strength of the phase transition as shown in figure~\ref{GW-Spectrum}.
In figure~\ref{DetectableRegion}(right), we show the correlations between $p_{runaway}$ and $\frac{v_n}{T_n}$. For
a successful EWBG the primordial phase transition must be strong first-order, i.e. $\frac{v_n}{T_n}>1$, the
bubble wall must not runaway, i.e. $p_{runaway}< 0$. Requiring these reveals that for an inert singlet model there
is an upper bound on the strength of the first-order EWPT, $\frac{v_n}{T_n} < 5$.

It will be interesting to assess the persistence of this result within other models such as those based on
supersymmetry~\cite{AbdusSalam:2011fc, AbdusSalam:2013qba, AbdusSalam:2014uea, AbdusSalam:2015uba,
  AbdusSalam:2017uzr, AbdusSalam:2019gnh} and its possible interplay, complementarity or synergy to collider and dark matter 
phenomenology~\cite{AbdusSalam:2019kei}. The same applies to other beyond the SM (BSM) extensions such as the Higgs sector-only
BSMs (see e.g.~\cite{AbdusSalam:2013eya, YaserAyazi:2019caf, Liu:2017gfg, Blinov:2015vma, Bell:2020gug}), the effective field
theory framework~\cite{Dawson:2017vgm} and exotics~\cite{Blinov:2015sna}. It would also be interesting to analyse the wall
velocity~\cite{Moore:1995ua, Moore:1995si, Moore:2000wx, John:2000zq, Megevand:2009ut, Megevand:2009gh, Dorsch:2018pat}
variations with respect to model parameter space such as for labelling regions with different possible bubble
expansion types -- detonations, deflagrations, or hybrid -- and 
corresponding correlations with GW strength~\cite{Ellis:2018mja, Ellis:2019oqb, Zhou:2019uzq, Dev:2019njv}. We hope to address some of these topics in
the near future. 

\begin{figure}
{\begin{tikzpicture}
\draw (.4,0) node[above right]{\includegraphics[width=0.45\linewidth]{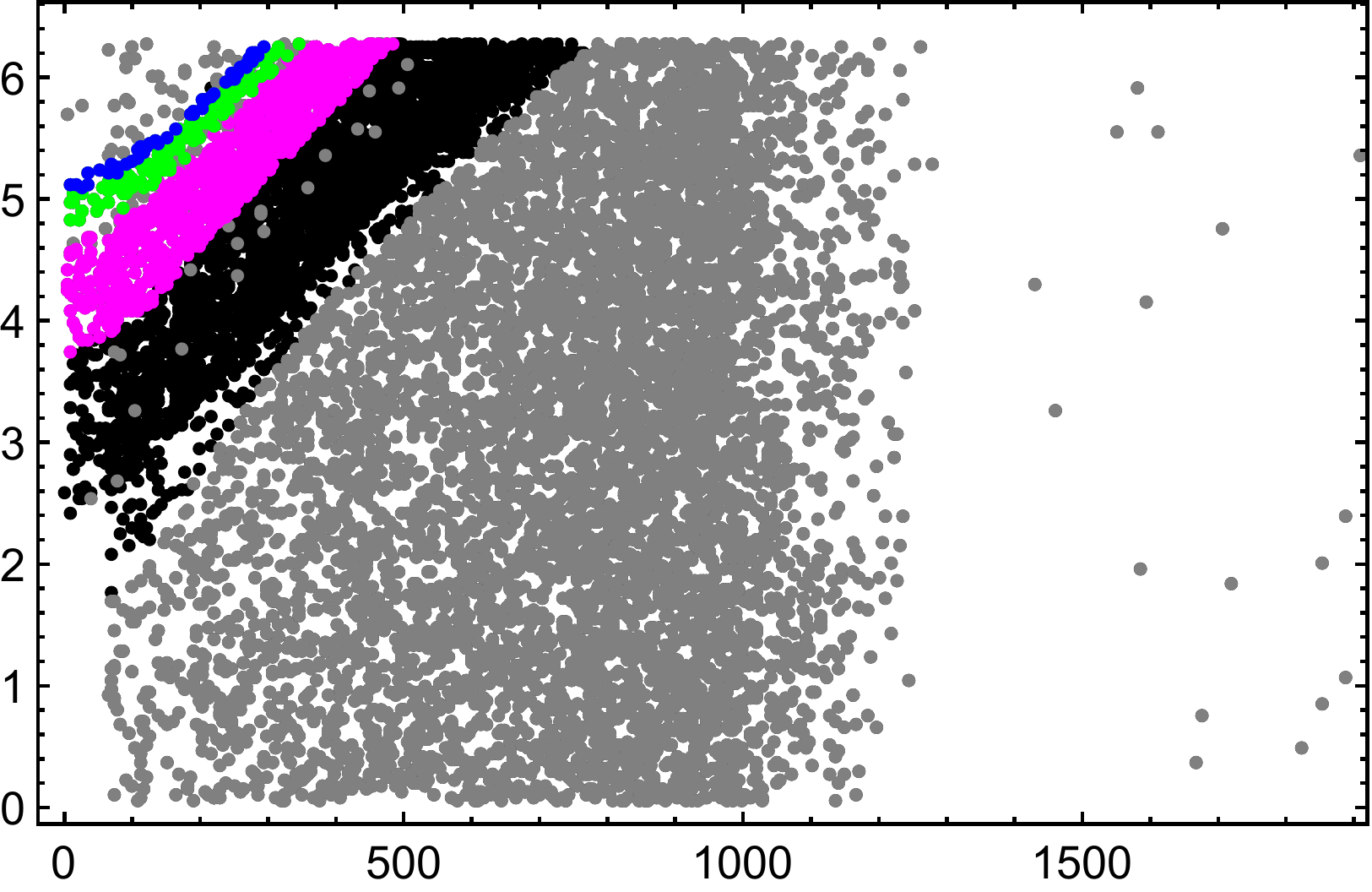}};
\draw (7.2,0) node[above right]{\includegraphics[width=0.47\linewidth]{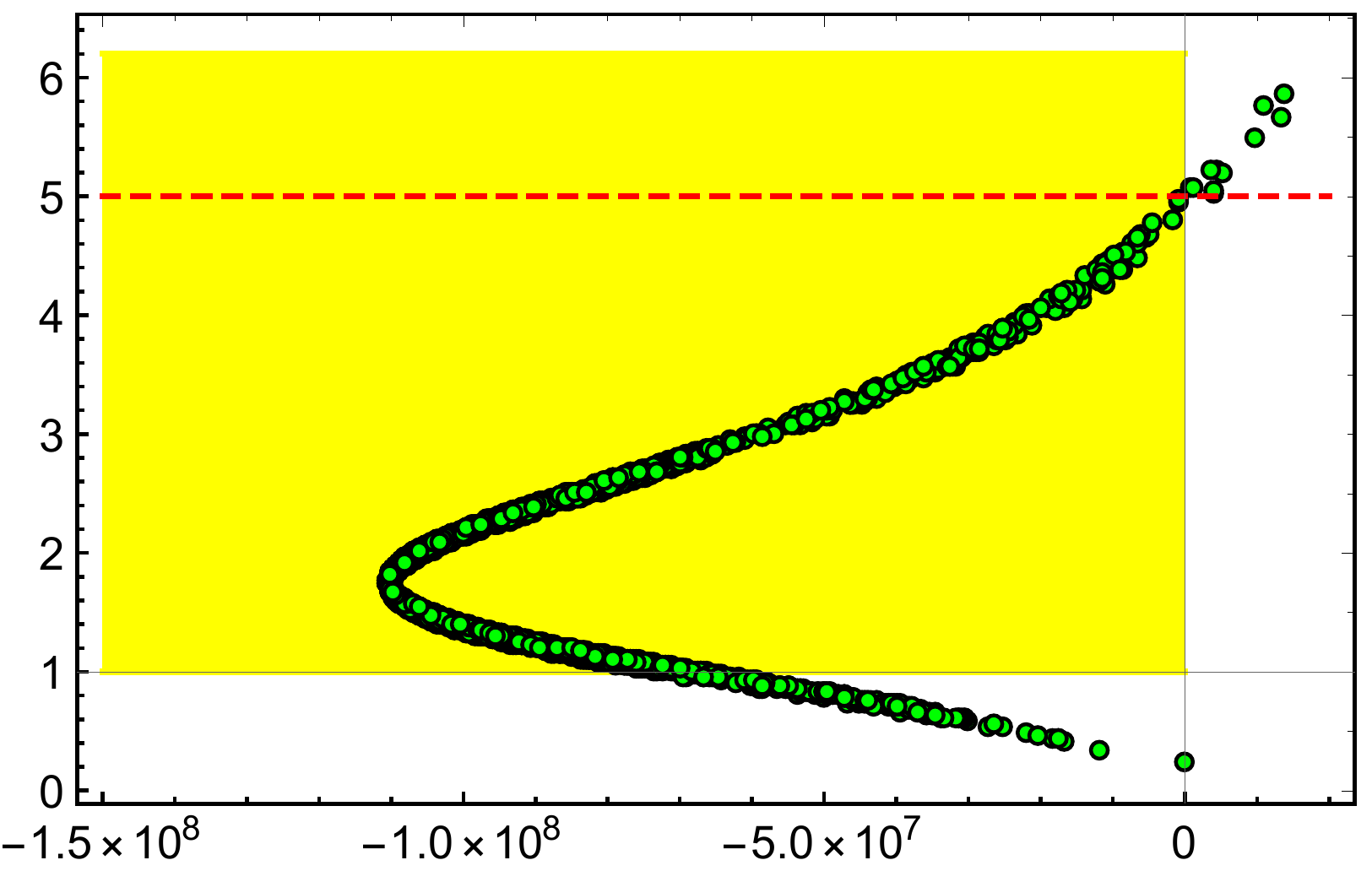}};
\draw (7.0,2.2) node[above right]{\large{$\frac{v_n}{T_n}$}};
\draw (-.2,2.2) node[above right]{$\lambda_{SH}$};
\draw (3,-.5) node[above right]{$\mu_S\rm{(GeV)}$};
\draw (10,-.5) node[above right]{$p_{runaway}\rm{(GeV^4)}$};
\end{tikzpicture}
}
  \caption{\label{DetectableRegion}  {\bf (Left)} The Scatter plot of $\mu_S$ and $\lambda_{HS}$ parameters. Gray points do not lead to first order EWPT, black points leads to first order EWPT but not detectable GW and other points lead to detectable GW using future space-based GW detectors; Lisa (blue), BBO (magenta) and DECIGO (green).
    {\bf (Right)} The Scatter plot of $p_{\rm runaway}$ versus $\frac{v_n}{T_n}$. For a successful EWBG, the following
    conditions must be satisfied: (i) the phase transition must be strongly first order, i.e. $\frac{v_n}{T_n}>1$,
    and (ii) The bubble wall must not runaway, i.e. $p_{\rm runaway}< 0$. This result shows that for the inert singlet model,
    the second condition is equivalent to  $\frac{v_n}{T_n}  \lesssim  5$;  an EWPT with $\frac{v_n}{T_n} > 5$    leads to a runaway bubble wall scenario.}
\end{figure}
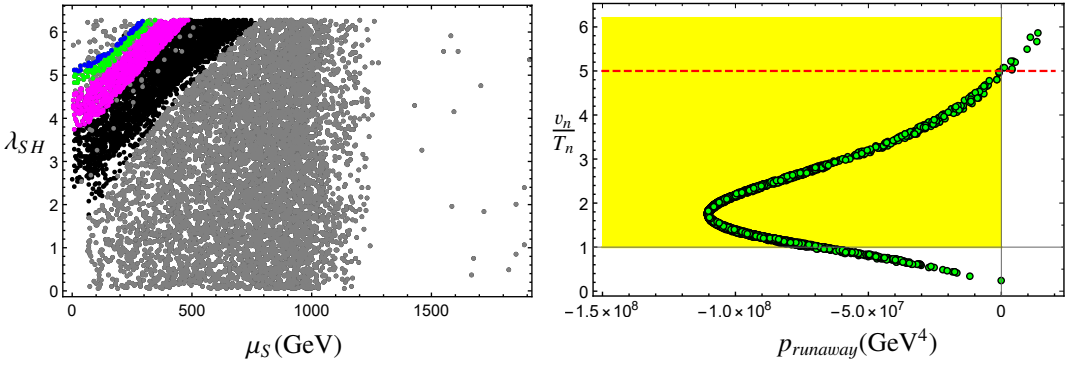

\paragraph{Acknowledgements:} Thanks very much to C.L. Wainwright for his attention concerning cosmoTransitions
package and to S. Sadat-Gousheh, Mohammad Mohammadi-Doust, Safura Sadeghi, Mehrdokht Sasanpour, H. Hashamipour,
A. Kargaran, Ankit Beniwal, Marek Lewicki, and Philipp Basler for conversations, useful discussions, or 
comments while working on this project.

\appendix
\section{Effective potential} \label{Veff}
The inert singlet model effective potential for our work is based on~\cite{Quiros:1999jp}. It is composed of the
tree level terms, $V_{tree}(h,S)$, the Coleman-Weinberg zero-temperature quantum correction terms, $V_{1-loop}(h,S)$,
and thermal correction terms, $V_T(h,S,T)$ \cite{Dolan:1973qd,Weinberg:1974hy}:
\begin{equation}
V_{eff}(h,S,T) = V_{tree}(h,S)+V_{1-loop}(h,S)+V_T(h,S,T). 
\end{equation}
The zero temperature one-loop correction~\cite{Quiros:1999jp,Curtin:2014jma} in the on-shell renormalisation scheme
with cutoff regularisation is given by: 
\begin{equation}
  V_{1-loop}(h,S) = \sum_{h,\chi,W,Z,t,S} \frac{n_i}{64 \pi^2} \left[m_i^4(h,S) \,
    \left( \log \frac{m_i^2(h,S)}{m_{0i}^2(v,0)} - \frac{3}{2}    \right)
    + 2m_i^2(h,S) \, m_{0i}^2(v,0)
    \right]  
\end{equation}
where $n_{i = h,\chi,W,Z,t,S} = \{1, 3, 6, 3, -12, 1\}$ and $m_0(v,0)$ are masses calculated at the electroweak VEV $S = 0, h = v$. The field dependant masses are:
\begin{equation}
  m_W^2 = \frac{g^2}{4}h^2, \quad \quad m_z^2 = \frac{g^2+g^{\prime 2}}{4}h^2, \quad \quad
  m_t^2 = \frac{y_t^2}{2}h^2, \quad \quad m_{\chi}^2 = -\mu^2+\lambda_Hh^2+\lambda_{HS}S^2.
\end{equation}
For the Higgs and the inert scalar singlet, the field-dependent masses are the eigenvalues of
the h and S mass mixing matrix:
\begin{equation}
  M^2_{HS}=  
 \begin{pmatrix}
   -\mu^2+3 \lambda_Hh^2+ \lambda_{HS}S^2 & 2 \lambda_{HS}hS \\
   2 \lambda_{HS}hS & \mu_S^2+3 \lambda_S S^2+ \lambda_{HS}h^2
 \end{pmatrix}. 
\end{equation}
Lastly, the thermal correction terms in the inert singlet effective potential are~\cite{Quiros:1999jp,Curtin:2014jma}:
\begin{equation}
  V_T(h,S,T) =  \sum_{h,\chi,W,Z,t,S} \frac{n_i T^4}{2\pi^2}\, J_b\left( \frac{m_i^2(h,S)}{T^2} \right)
  + \sum_{i=t} \frac{n_i T^2}{2 \pi^2} \, J_f\left( \frac{m_i^2(h,S)}{T^2} \right)
\end{equation}  
where  
\begin{equation}
J_{b/f}(x) = \int_0^{\infty} \, dk \, k^2 \, \log \left[ 1 \mp \exp \left(\sqrt{k^2 + x} \, \right) \right]. 
\end{equation}  
Expand to leading order in $(\frac{m}{T})^2$, the thermal corrections to the scalar masses in the inert
singlet model are can be determined as the eigenvalues of the mass matrix 
\begin{equation}
  M_{HS}^2+
   \begin{pmatrix}
   \Pi_h(T) & 0 \\
   0 & \Pi_S(T)
 \end{pmatrix}
\end{equation}
where 
\begin{equation}
  \Pi_h(T) = \Pi_{\chi}(T) = T^2(\frac{g^{\prime 2}}{16} +
  \frac{3g^2}{16} + \frac{\lambda_H}{2}+\frac{y_t^2}{4}+\frac{\lambda_{SH}}{12}), \quad  
  \Pi_s(T) = T^2(\frac{\lambda_{HS}}{3}+\frac{\lambda_S}{4}),
  \textrm{ and } \Pi_W(T)=\frac{11}{6}g^2T^2. 
\end{equation}  
The corrected masses of Z-boson and and photon, $\gamma$, are the eigenvalues of the mass matrix 
\begin{equation}
   \begin{pmatrix}
   \frac{1}{4}g^2h^2+\frac{11}{6}g^2T^2 & -\frac{1}{4}g^{\prime}gh^2 \\
   -\frac{1}{4}g^{\prime}gh^2 & \frac{1}{4}g^{\prime 2}h^2+\frac{11}{6}g^{\prime 2}T^2
 \end{pmatrix}. 
\end{equation}

\section{Inert Real Triplet Model} \label{IRT}
For comparison, we also analyse another model beyond the SM, constructed via extending the Higgs sector by adding a triplet representation of SU(2) group -- the inert real triplet model (IRTM). In this model, the tree level potential is given by
\begin{equation}
V(H,\Delta)=-\mu_H^2 |H|^2 + \lambda_H |H|^4+\frac{1}{2}\mu_\Delta^2 |\Delta|^2 + \frac{1}{4}\lambda_\Delta |\Delta|^4+\lambda_{H\Delta} |H|^2|\Delta|^2
\end{equation}
where
$H=\frac{1}{\sqrt{2}}(G^+,h+iG^0)^T,$
and
$\Delta=(\Delta^+,\Delta^0,\Delta^-)^T$. It leads to
\begin{equation}
V_{tree}(h,\sigma)=-\frac{1}{2}\mu_H^2  h^2 + \frac{1}{4}\lambda_H  h^4+\frac{1}{2}\mu_\Delta^2  \sigma^2 + \frac{1}{4}\lambda_\Delta  \sigma^4+ \frac{1}{2} \lambda_{H\Delta} h^2\sigma^2
\end{equation}
where
$\sigma=\langle \Delta^0 \rangle$.

The zero-temperature and finite-temperature corrections, at one-loop level, are given by 
\begin{equation}
  V_{1-loop}(h,\sigma) = \sum_{h,G,W,Z,t,\Delta^{0,\pm}} \frac{n_i}{64 \pi^2} \left[m_i^4(h,\sigma) \,
    \left( \log \frac{m_i^2(h,\sigma)}{m_{0i}^2(v,0)} - \frac{3}{2}    \right)
    + 2m_i^2(h,\sigma) \, m_{0i}^2(v,0)
    \right]  
\end{equation}
\begin{equation}
  V_T(h,\sigma,T) =  \sum_{h,G,W,Z,t,\Delta^{0,\pm}} \frac{n_i T^4}{2\pi^2}\, J_b\left( \frac{m_i^2(h,\sigma)}{T^2} \right)
  + \sum_{i=t} \frac{n_i T^2}{2 \pi^2} \, J_f\left( \frac{m_i^2(h,\sigma)}{T^2} \right).
\end{equation}  
Here the field-dependent masses are given by
$$m_W^2=\frac{g^2}{4}(h^2+3\sigma^2),\ \ \ \ m_Z^2=\frac{1}{4}(g^2+g'^2)h^2, \ \ \ \  m_t^2=\frac{y_t^2}{2}h^2,$$
$$m_{G^{0,\pm}}^2=-\mu_H^2+\lambda_H h^2+\lambda_{H\Delta}\sigma^2, \ \ \ \ m_{\Delta^\pm}^2=\mu_\Delta^2+\lambda_{H\Delta}h^2+\lambda_\Delta\sigma^2.$$
The mass-squares of the $h$ and $\Delta^0$ fields are given by the eigenvalues of the matrix
\begin{equation}
  M^2_{h\sigma}=  
 \begin{pmatrix}
   -\mu_H^2+3 \lambda_H h^2+ \lambda_{H\Delta}\sigma^2 & 2 \lambda_{H\Delta} h\sigma \\
   2 \lambda_{H \Delta} h\sigma & \mu_\Delta^2+3 \lambda_\Delta \sigma^2+ \lambda_{H\Delta}h^2
 \end{pmatrix}. 
\end{equation}
We randomly scan the parameter space, in the region: 
$$10 (GeV)<\mu_\Delta<1000(GeV),$$  
$$0.1<\lambda_{H\Delta}<10,$$
$$0.1<\lambda_{\Delta}<1.$$ 
The results, as shown in the figures that follow are similar to those for the inert singlet model. There are strong correlation between $p_{\rm runaway}$ and $\frac{v_n}{T_n}$.

\begin{figure}[H]
{\begin{tikzpicture} 
\draw (3,0) node[above right]{\includegraphics[width=0.7\linewidth]{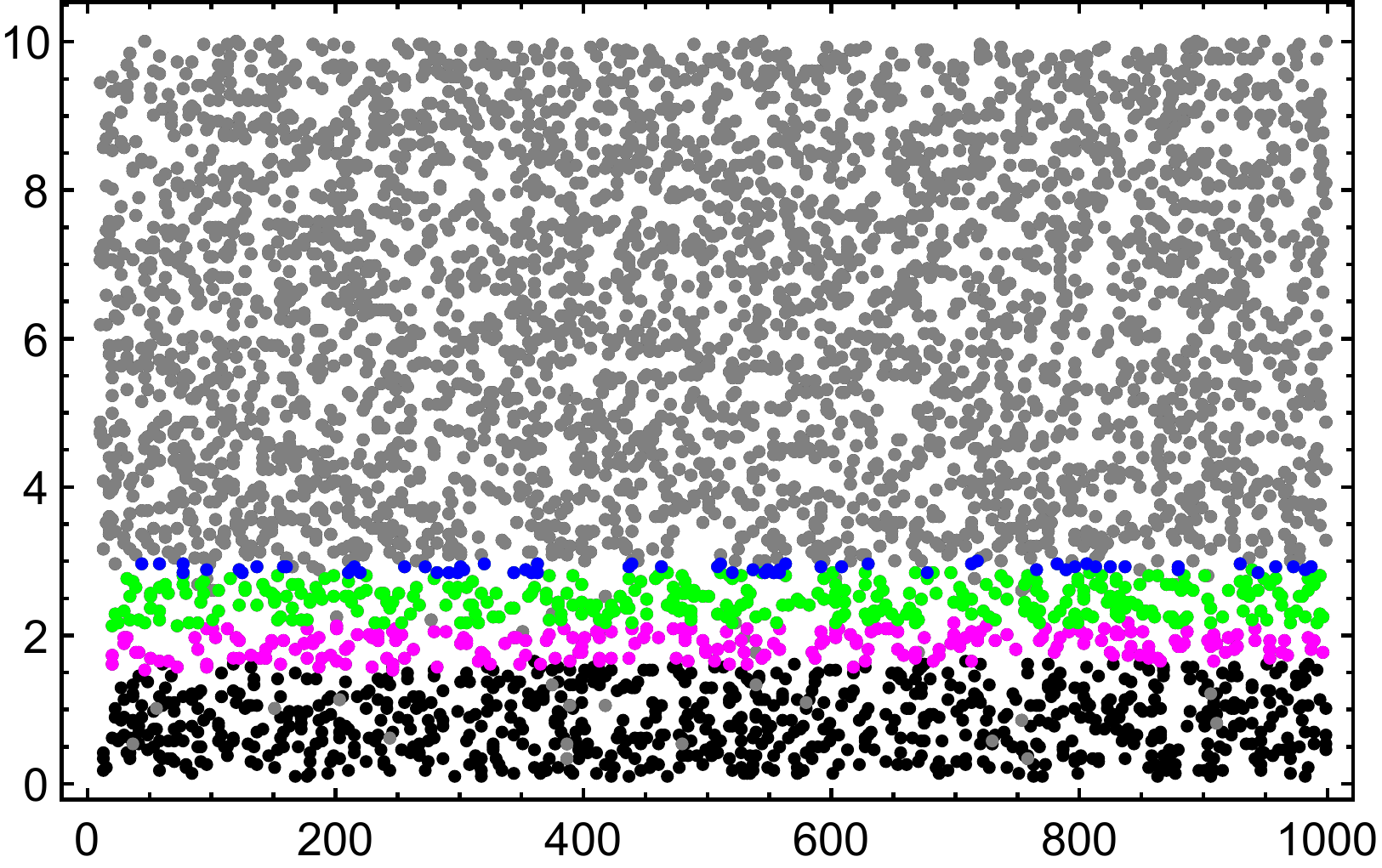}};
\draw (2,3) node[above right]{$\lambda_{H\Delta}$};
\draw (7.3,-.5) node[above right]{$\mu_\Delta\rm{(GeV)}$};
\end{tikzpicture}
}
\caption{\label{DetectableRegion}  The Scatter plot of $\mu_S$ and $\lambda_{H\Delta}$ parameters. Gray points do not lead to first order EWPT, black points leads to first order EWPT but not detectable GW and other points lead to detectable GW using future space-based GW detectors; Lisa (blue), BBO (magenta) and DECIGO (green). }
\end{figure}
\begin{figure}[H]
{\begin{tikzpicture} 
\draw (3,0) node[above right]{\includegraphics[width=0.7\linewidth]{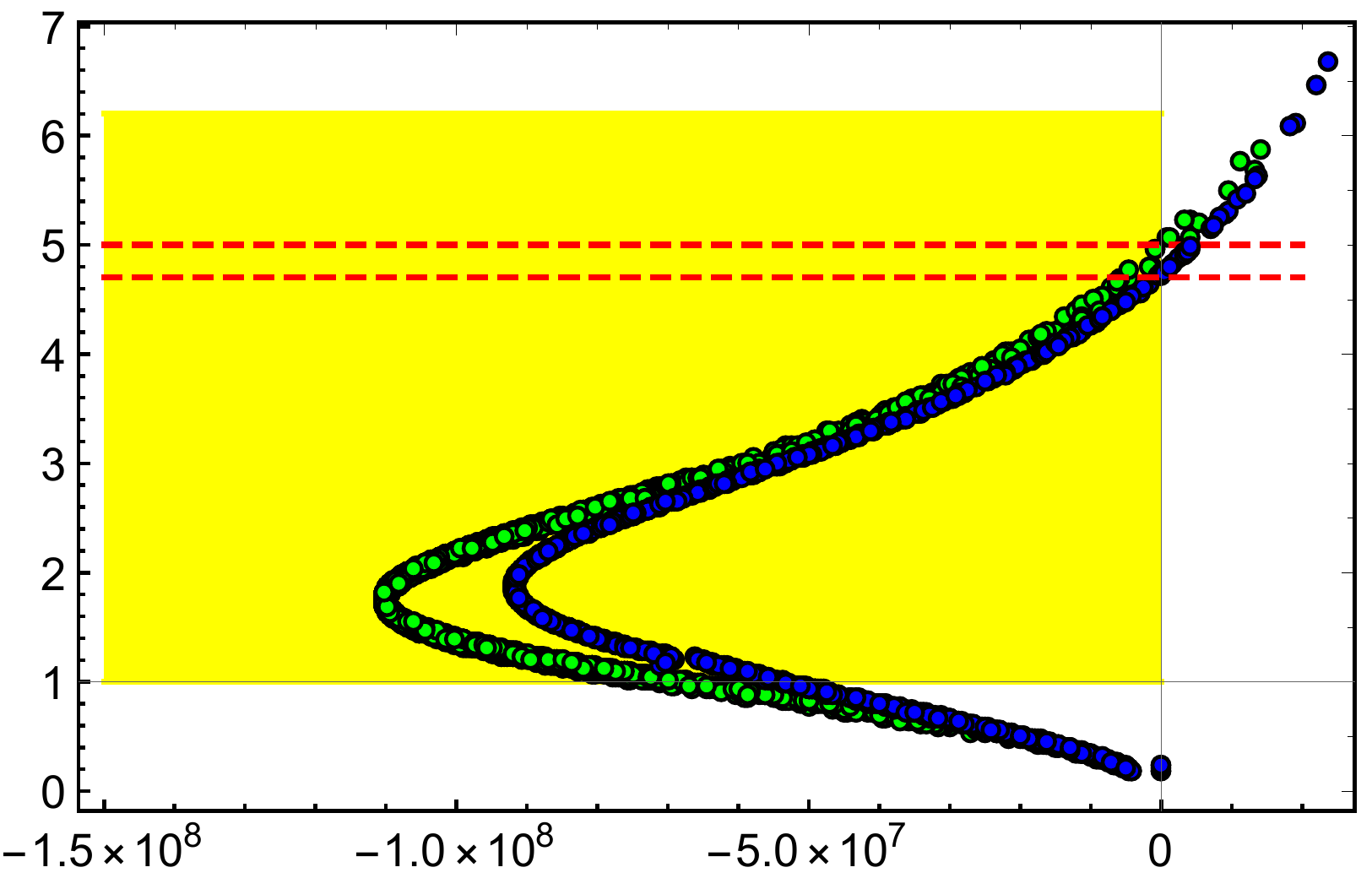}};
\draw (2.5,3) node[above right]{{\large{$\frac{v_n}{T_n}$}}};
\draw (7.3,-.5) node[above right]{$p_{runaway}\rm{(GeV^4)}$};
\end{tikzpicture}
}
\caption{\label{DetectableRegion2}  The Scatter plot of $p_{\rm runaway}$ versus $\frac{v_n}{T_n}$. For a successful EWBG, the following conditions must be satisfied: (i) the phase transition must be strongly first order, i.e. $\frac{v_n}{T_n}>1$, and (ii) The bubble wall must not runaway, i.e. $p_{\rm runaway}< 0$. This result shows that  the second condition is equivalent to  $\frac{v_n}{T_n}  \leq 5$ for the inert real singlet model (green), and  $\frac{v_n}{T_n}  \leq 4.7$ for the inert real triplet model (blue).}
\end{figure}
\begin{figure}[H]
{\begin{tikzpicture} 
\draw (0,0) node[above right]{\includegraphics[width=0.7\linewidth]{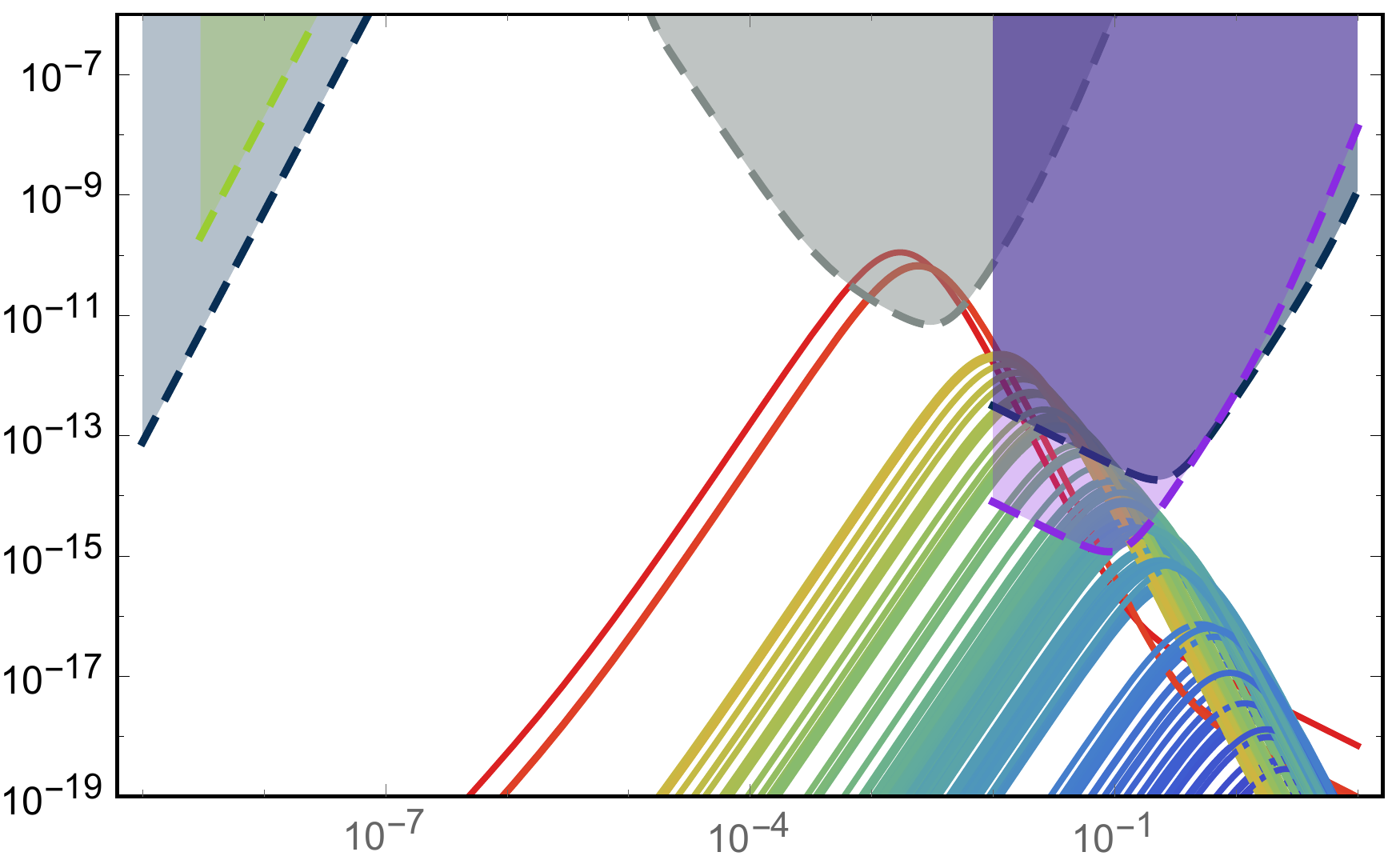}};
\draw (9.5,-0.1) node[above right]{\includegraphics[width=0.062\linewidth]{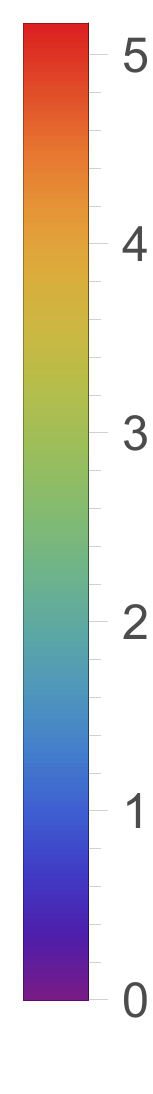}};
\draw (9.6,6) node[above right]{\Large {$\frac{v_n}{T_n}$}};
\draw (4.5,-.6) node[above right]{$f(\rm{Hz})$};
\draw (-1.4,3.) node[above right]{$\Omega h^2(f)$};
\draw (1.,5) node[above right]{$\rm{EPTA}$};
\draw (1.,3.5) node[above right]{$\rm{SKA}$};
\draw (5.4,5) node[above right]{$\rm{LISA}$};
\draw (7.2,5) node[above right]{$\rm{DECIGO}$};
\draw (7.0,2.2) node[above right]{$\rm{BBO}$};
\end{tikzpicture}
}
\caption{  Spectra of GWs from the electroweak phase transition for randomly
  sampled examples from the coloured points in figure \ref{DetectableRegion2}, i.e. the points with
  strong first-order EWPT. The sensitivity region for prospective GW detectors such as LISA, BBO and
  DECIGO are also shown. It can be seen that the intensity of GW signal increases with the strength
  of the phase transition, i.e. $\frac{v_n}{T_n}$. For comparison we also show the sensitivity
  regions for SKA and EPTA detectors which cannot probe any part of the inert singlet parameter space.} 
\end{figure}

\end{document}